\documentclass[mathleft
]{an}
\usepackage{graphicx}
\usepackage{times}
\usepackage{natbib}
\begin{document}

\Pagespan{1}{}
\Yearpublication{2010}%
\Yearsubmission{2010}%
\Month{11}%
\Volume{999}%
\Issue{88}%

\title{New observations of ULX supershells, and their implications}

\author{David M. Russell\inst{1}\fnmsep\thanks{Corresponding author:
  \email{d.m.russell@uva.nl}\newline},
Yi-Jung Yang\inst{1}, Jeanette C. Gladstone\inst{2}, Klaas Wiersema\inst{3} \and Timothy P. Roberts\inst{4}
}
\titlerunning{ULX supershells}
\authorrunning{Russell, Yang, Gladstone, Wiersema \& Roberts}
\institute{
Astronomical Institute `Anton Pannekoek', University of Amsterdam, PO Box 94249, 1090 GE Amsterdam, the Netherlands
\and 
Department of Physics, University of Alberta, 11322-89 Avenue, Edmonton, AB T6G 2G7, Canada
\and 
Department of Physics \& Astronomy, University of Leicester, University Road, Leicester, LE1 7RH, UK
\and 
Department of Physics, University of Durham, South Road, Durham, DH1 3LE, UK
}

\received{}
\accepted{}
\publonline{later}


\abstract{%
New optical narrowband imaging observations of the fields of several ULXs are presented. Known supershell nebulae are associated with a number of these ULXs, which we detect in emission lines filters such as [S II], He II, [O II] and [O III]. New nebulae are discovered, which are candidate ULX-powered supershells. The morphologies and emission line fluxes of these nebulae could then be used to infer the properties of the emitting gas, which gives clues to the energizing source (photoionization and/or shock-excitation, both possibly from the ULX). Studies of supershells powered by ULXs can help to constrain the nature of ULXs themselves, such as the isotropy of the X-ray emission and the strength of their outflows.
}

\maketitle

\def\simlt{\mathrel{\rlap{\lower 3pt\hbox{$\sim$}}
        \raise 2.0pt\hbox{$<$}}}
\def\simgt{\mathrel{\rlap{\lower 3pt\hbox{$\sim$}}
        \raise 2.0pt\hbox{$>$}}}

\bibpunct{(}{)}{;}{a}{}{,}

\section{Introduction}
Ultraluminous X-ray sources (ULXs) are objects emitting at X-ray luminosities exceeding $\sim 10^{39}$ erg s$^{-1}$ (assuming the radiation is isotropic) that usually reside within galaxies but not at their centres. They are not considered to be supermassive black holes (e.g. active galactic nuclei; AGN), nor can they be stellar-mass compact objects (black holes or neutron stars) accreting at sub-Eddington rates from a companion star (X-ray binaries; XBs). Their nature remains generally elusive, although most are likely to be XBs harbouring black holes of tens (but not hundreds) of solar masses (Jin, Feng \& Kaaret 2010) accreting in specific states which can appear super-Eddington (Gladstone, Roberts \& Done 2009). Some of the most luminous ULXs (termed hyperluminous X-ray sources) may contain intermediate-mass black holes (hundreds of solar masses; e.g. Farrell et al. 2009).

Unlike most XBs, but like many AGN, ULXs are often found to be associated with large, resolved structures. These are ULX nebulae (see e.g. Pakull \& Gris\'e 2008), ionized gas surrounding the ULX, typically found in optical line emission (many ULXs are also in proximity to star-forming regions; Swartz, Tennant \& Soria 2009). Since ULXs are by definition strong X-ray/UV emitters, their radiation photoionizes the local interstellar medium (ISM), producing observable nebulae tens of parsecs in size (e.g. the He III region surrounding the ULX in Holmberg II; Kaaret, Ward, \& Zezas 2004). Such nebulae can be used to infer the \emph{isotropic} luminosity of the ULX, and hence test for relativistic beaming of its X-rays. The larger scale analogies are AGN, which heat and ionize the intergalactic medium, suppressing cooling flows and galaxy formation. On smaller scales, an X-ray ionized nebula of radius $\sim 5$ pc is found associated with the black hole XB LMC X--1 (e.g. Cooke et al. 2008).

Evidence is mounting that ULXs also release energy kinetically via powerful, collimated jets/outflows, as is the case for both XBs and AGN. While the jets themselves are unlikely to be detected because their expected radio fluxes are fainter than current radio telescope detection limits 
\newline
 (K\"ording, Colbert \& Falcke 2005), ULX jets and/or winds have been inferred energetically via `superbubbles' hundreds of parsecs in diameter and some bipolar in morphology, which are powered collisionally (e.g. Ramsey et al. 2006; Abolmasov et al. 2007; Abolmasov 2010). Historically, radio lobes in FRII radio galaxies are used as calorimeters to infer the kinetic power contained in the AGN jets fueling them. In the last few years these techniques have been successfully applied to the few known examples of XB jet--ISM interactions (e.g. Gallo et al. 2005; Pakull, Soria \& Motch 2010). Quite surprisingly, the results so far suggest that a large fraction of the accretion energy is channeled into the mechanical power in the jets (Dubner et al. 1998; Russell et al. 2007; Cooke et al. 2007; Sell et al. 2010; Soria et al. 2010); a result that is a large step forward in our understanding of accretion physics in strong gravitational fields. The jet powers cannot be inferred from observations of the jets themselves since an uncertain fraction of the energy in the jets is radiated away (perhaps $\sim 5$\%; Fender 2001).

Unlike AGN, few large-scale jet-powered lobes/bubbles have been discovered so far associated with XBs (e.g. Mirabel et al. 1992; Miller-Jones et al. 2008), and indeed more candidates may exist for ULXs. For XB jets, a high local mass density is required for a bow shock to form (Heinz 2002), the surface brightness of the radio lobes are typically below current detection limits (Kaiser et al. 2004), and a low space velocity of the system is needed so that the power is not dissipated over too large a volume (although such structures could still be detected in some cases; Heinz et al. 2008; Wiersema et al. 2009). If the ratio between jet power and X-ray luminosity could be similar for ULXs and XBs, ULXs may be energetically more favourable for detections of jet--ISM interactions. In addition, their large extent of hundreds of parsecs would be hard to detect in our own Galaxy because the nebulae would be too diffuse, spanning several degrees and requiring very wide field imaging.

\begin{table}
\caption{Filters used. $\lambda$ is the central wavelength. The redshifted H$\alpha$ filter is denoted by H$\alpha$z.}
\begin{tabular}{lrrrr}\hline
Filter & \multicolumn{2}{c}{------- WFC -------} & \multicolumn{2}{c}{------ DFOSC ------} \\
       & $\lambda$ ($\rm \AA$) & FWHM ($\rm \AA$) & $\lambda$ ($\rm \AA$) & FWHM ($\rm \AA$) \\ 
\hline
~[O II]   & 3727 & 100 & & \\
~He II     & 4686 & 100 & & \\
~H$\beta$  & 4861 & 170 & & \\
~[O III]  & 5008 & 100 & 5010 & 57 \\
~Harris V  & 5425 & 975 & & \\
~H$\alpha$ & 6568 & 95 & 6562 & 62 \\
~H$\alpha$z& 6657 & 79 & &  \\
~[S II]   & 6725 & 80 & 6727 & 63 \\
\hline
\end{tabular}
\end{table}

Observations of ULX nebulae can therefore provide a powerful tool to help uncover the nature of ULXs; placing constraints on their isotropic X-ray/UV luminosity and mechanical output. Here we present optical narrowband observations of the fields of several ULXs and identify a number of candidate ULX-powered nebulae.

\section{Observations}

Imaging of six galaxies containing a total of 17 ULXs (Liu \& Mirabel 2005) were performed using two telescopes. We observed on three consecutive nights in April 2009 using the Wide Field Camera (WFC) on the 2.5-m Isaac Newton Telescope (INT) located at La Palma, Spain. The WFC has a field of view (FOV) of 34$\times$34 arcmin. One target, NGC 7793, was observed in September 2006 with the Danish 1.5-m telescope at La Silla, Chile, equipped with the Danish Faint Object Spectrograph and Camera (DFOSC; with a FOV of 13.7 $\times$ 13.7 arcmin). Some ULXs have known nebulae and some do not; for those which do we used many narrowband filters because the fluxes and emission line ratios can be used to infer the properties of the emitting gas. For instance, the [S II] doublet is most luminous for shock-ionized gas and the He II line is strongest for photoionized gas. For the WFC observations a redshifted narrowband H$\alpha$ filter was used for continuum subtraction for filters centred in the red (the same exposure times were used); for filters centred at blue wavelengths the broadband Harris V filter was used (with a shorter exposure time). By subtracting the continuum emission, most of the galaxy light is removed and just the line emission is revealed. The filters used are listed in Table 1.

In addition, spectra of the optical counterpart of NGC 5204 X--1 were obtained with the 8.1-m Gemini North telescope using the Gemini Multi-Object Spectrograph (GMOS) in July 2008. The slit used was 0.75 arcsec in width and 330 arcsec in length, orientated north--south. The log of all observations is presented in Table 2.

\begin{table*}
\caption{Log of observations. For filter wavelengths see Table 1. In the fourth column, 4$\times$10 refers to 4 exposures of 10 minutes each.}
\begin{tabular}{lllll}\hline
Date (UT) & Instrument & Galaxy & Filters (exposure times in minutes) & Airmass\\ 
\hline
2006 September 1 & DFOSC & NGC 7793 & H$\alpha$ (4$\times$10), [O III] (3$\times$10), [S II] (3$\times$10) & 1.00--1.18 \\
2008 July 3 & GMOS & NGC 5204 & Spectra: wavelength range 3800--6000 $\rm \AA$ (4$\times$12) & $\sim 1.4$ \\
2009 April 9 & WFC & NGC 4861 & H$\alpha$ (20+2), [S II] (20), H$\alpha$z (20) & 1.01--1.04 \\
2009 April 9 & WFC & NGC 4559 & H$\alpha$ (20), [S II] (20), H$\alpha$z (20) & 1.14--1.27 \\
2009 April 10 & WFC & NGC 5204 & H$\alpha$ (10), H$\alpha$z (10), [S II] (10), V (2.5), [O II] (10), H$\beta$ (10) & 1.20--1.30 \\
2009 April 10 & WFC & NGC 4395 & H$\alpha$ (20), H$\alpha$z (20), [S II] (20) & 1.01--1.04 \\
2009 April 11 & WFC & NGC 4559 & [O III] (20), [O II] (20), He II (20) & 1.00--1.03 \\
2009 April 11 & WFC & NGC 5204 & [S II] (10), [O III] (10), He II (10), H$\alpha$z (10) & 1.18--1.22 \\
2009 April 11 & WFC & NGC 5194 & [S II] (10), H$\alpha$z (10) & 1.16--1.18 \\
\hline
\end{tabular}
\end{table*}

Data reduction of the imaging data was performed using the pipeline package \small THELI \normalsize (Erben et al. 2005) for the WFC data and \small IRAF \normalsize for the DFOSC data. The science images were de-biased and flat-fielded using master bias and master flat frames. Positional calibration of the WFC data was achieved by matching several hundred stars in each exposure with those in the online MAST Guide Star Catalog (release 2.3) using \small THELI\normalsize. The DFOSC images were aligned and stacked in \small IRAF \normalsize to produce one deep image per filter. The GMOS spectra were reduced (de-biased, flat-fielded, sky subtracted, combined, flux calibrated and wavelength calibrated) using standard Gemini packages within \small IRAF \normalsize.

\begin{figure}
\centering
\includegraphics[width=83mm]{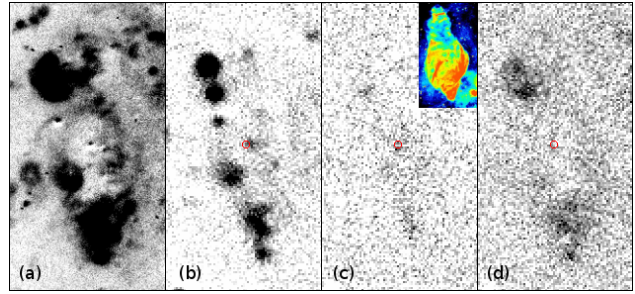}
\caption{Continuum-subtracted images of the region around NGC 5204 X-1 (north is up, east to the left). (a): H$\alpha$ (Canada France Hawaii Telescope image; R-band is used for continuum subtraction; data first presented in Pakull \& Mirioni 2002). (b--d): WFC images in [O III], He II and [S II], respectively. The FOV of each panel is 48 $\times$ 26 arcsec ($\sim 1120 \times 610$ pc at a distance of 4.8 Mpc). The Chandra error circle of the position of the ULX (Goad et al. 2002) is shown in red. The inset in (c) is a radio image of the W50 nebula distorted by the jets of SS433 (Dubner et al. 1998), at the same size scale (assuming a distance of 5.5 kpc to SS433 and W50; Lockman, Blundell \& Goss 2007) showing that the NGC 5204 X-1 nebula is $\sim 2$--3 times larger in length than W50.}
\end{figure}

\section{Results}

X-ray ionized nebulae are at most tens of parsecs from the ULXs energizing them and hence may be unresolved in our data (one example being NGC 5408 X--1; Kaaret \& Corbel 2009). However, ULX superbubbles that are energized collisionally are typically much larger, easily resolvable at these distances and may even appear detached from the ULXs since the evolution of jets causes underdense cavities surrounding XBs to develop (e.g. Hao \& Zhang 2009). We do indeed find nebulae in proximity to several ULXs, some already reported and some new. Here we present some interesting preliminary results from this study. In future work, we will use the emission line fluxes of the nebulae to infer their nature, estimate the photon and/or mechanical energy input and assess whether they could be powered by ULXs.

NGC 5204 X--1 is a ULX with a variable X-ray luminosity of $\sim 1$--6 $\times 10^{39}$ erg s$^{-1}$ (0.5--8 keV) residing in a warped spiral galaxy at a distance of 4.8 Mpc (e.g. Roberts et al. 2006). Roberts et al. (2001) detected ionized gas within 100 pc of the ULX in H$\alpha$, H$\beta$, [S II] and [O III], with an apparent cavity on one side; Abolmasov et al. (2007) found several emission lines in the optical spectra; and a 360 pc diameter circular `bubble' with an even larger, apparently bipolar structure was seen in an H$\alpha$ image by Pakull \& Mirioni (2002). In Fig. 1 we present some of our continuum subtracted WFC images of the environment of this ULX (in Fig. 1a the high resolution H$\alpha$ image first presented in Pakull \& Mirioni 2002 is shown).
Regions of the circular bubble are detected in [O III] (Fig. 1b), suggesting that this may be a shock wave travelling at $\simgt 100$ km s$^{-1}$ (e.g. Allen et al. 2008). Abolmasov et al. (2007) also showed that a nebula close to the ULX could be energized collisionally and/or photoionized, from spectra. Since this nebula is part of the circular bubble, the morphology suggests a shock wave (which could also be partly photoionized). The two ionized regions to the north and south of the bubble are detected in [O III] (Fig. 1b), [S II] (Fig. 1d), [O II] and H$\beta$ (not shown) and faintly in He II (Fig. 1c).

Numerous emission lines from the nebula are visible in the GMOS spectrum of NGC 5204 X--1 (Gladstone 2009). The slit is orientated north--south, and emission from the ULX optical counterpart, the circular bubble and large scale bipolar structure are all detected. By comparing the de- reddened [O II] / H$\beta$ and [Ne III] / H$\beta$ emission line ratios with models for radiative shock waves (Allen et al. 2008), initial results suggest that much of the gas in the large-scale nebula is travelling at velocities of the order $\sim 200$ km s$^{-1}$. The non-detection of He II in the GMOS spectrum and only faint detection in some regions in the WFC image imply that photoionization does not play a large role in the energetics in most regions of the superbubble; the bright [O II] and [S~II] seen in the WFC images of the bipolar structure imply it could be collisionally energized. The brightest parts of the whole structure appear to form a straight line that almost passes through the position of the ULX (most visible here in the [O III] image). These preliminary results suggest this system could represent the first known ULX analogue of the large-scale SS433 / W50 complex (e.g. Dubner et al. 1998), in which the jets of SS433 are distorting and energizing the supernova remnant W50. W50 is 2--3 times smaller in length than the NGC 5204 X--1 superbubble (see inset in Fig. 1c); SS433 also has a lower X-ray luminosity.

NGC 4559 is a late-type spiral galaxy at a distance of $\sim 9.7$ Mpc (Sanders et al. 2003) that contains two ULXs; X10 ($L_{\rm X} \sim 1 \times 10^{40}$ erg s$^{-1}$; 0.3--10 keV) residing close to the centre of the galaxy (but not consistent with its nucleus) and X7 ($L_{\rm X} \sim 2$--$3 \times 10^{40}$ erg s$^{-1}$; 0.3--10 keV) in an isolated, outer region of the galaxy (Cropper et al. 2004). X7 is located within a large star-forming complex, studied extensively by Soria et al. (2005). The authors show that a ring-shaped H$\alpha$ structure is coincident with the star-forming regions and cannot be ionized (solely) by X-rays from the ULX. In Fig. 2 the fields of these two ULXs are shown in continuum-subtracted emission line images taken with the WFC. We find that the H$\alpha$ ring surrounding X7 is also bright in [O II] and [O III] (upper panels). Fainter detections are also made in [S II] and He II (not shown here). The line emission likely originates in gas ionized by the population of young massive stars, but we can use the emission line ratios to assess the possibility of photoionization or shock excitation from the ULX (in a follow-up work). An emission line nebula lies a few hundred parsecs north--east of the position of X10 (Fig. 2, lower panels). The western region of the nebula, closer to the ULX is brighter in [O II], [O III] and He II, whereas the eastern region is brighter in [S II]. This is a new candidate ULX superbubble, but further analysis is required to test whether this nebula could be energized by the ULX.

\section{Discussion and conclusions}

We have introduced a study into the ionized environments of a selection of ULXs, and give some preliminary results. Narrowband images were obtained using two telescopes in many emission line filters sensitive to photoionized and shock-ionized gas. Candidate ULX-powered supershell nebulae are identified, including known and new candidates. In a future work we will focus on individual ULXs and use the morphologies and line ratios to infer the likely energy input to these structures from the ULXs.

\begin{figure}
\includegraphics[width=83mm]{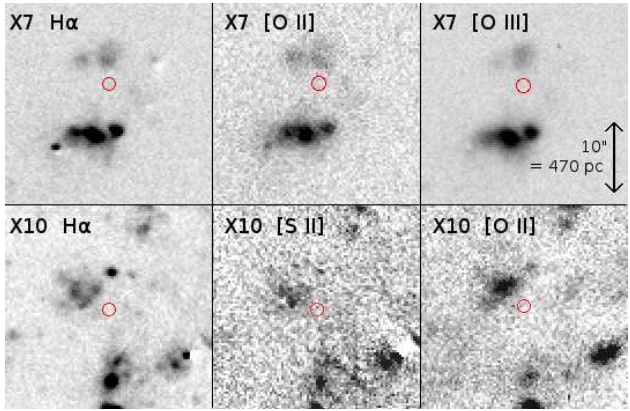}
\caption{Continuum-subtracted images of the region around NGC 4559 X7 (upper panels) and X10 (lower panels). Each panel is $30 \times 30$ arcsec ($\approx 1.4 \times 1.4$ kpc at a distance of 9.7 Mpc); north is up, east to the left. The red circles indicate the X-ray positions of the two ULXs (the error circles are smaller than the ones shown). We measure the positions from archival Chandra data using the source detection program \emph{celldetect} (Calderwood et al. 2001); these agree with the positions published in Cropper et al. (2004).}
\end{figure}

Although the environment of ULXs has only began to be studied in detail in the last decade, the number of regions in the environment likely energized, ionized or disrupted by ULXs has already exceeded the number of known cases for canonical XBs. This may be due to a combination of angular size scale issues (such diffuse structures in our galaxy are harder to identify since they would extend arcminutes or degrees across), lower extinction for extragalactic sources, the more luminous nature of ULXs compared to XBs and likely denser regions of the ISM around ULXs (which have young, high-mass companion stars) compared to low-mass XBs (which have older, low-mass stars). ULX nebulae may be abundant, and observable at many wavelengths with current facilities. Systematic, unbiased samples may prove pivotal in advancing our understanding of ULXs and the feedback they provide in heating the ISM in galaxies. The largest superbubbles appear to be at least partly jet--powered. For XBs, the kinetic energy of jets is hard to measure, but is a remarkably large fraction of the accretion energy in the few cases where estimates have been possible. ULXs may therefore provide the best opportunities to measure the energy contained in the jets produced by stellar-mass (and possibly even intermediate-mass) black holes.

\acknowledgements
The Isaac Newton Telescope is operated on the island of La Palma by the Isaac Newton Group in the Spanish Observatorio del Roque de los Muchachos of the Instituto de Astrofisica de Canarias. The Danish 1.54-m telescope is a Danish national facility operated by the European Southern Observatory. DMR acknowledges support from a Netherlands Organization for Scientific Research (NWO) Veni Fellowship.


\begin{thebibliography}{99}
  \bibitem[\protect\citeauthoryear{Abolmasov}{2010}]{abol10} Abolmasov, P.: 2010, New Astronomy, in press (arXiv:1007.4535)
  \bibitem[\protect\citeauthoryear{Abolmasov et al.}{2007}]{abolet07} Abolmasov, P., Fabrika, S., Sholukhova, O., Afanasiev, V.: 2007, AstBu, 62, 36
  \bibitem[\protect\citeauthoryear{Allen et al.}{2008}]{alleet08}Allen, M.G., Groves, B.A., Dopita, M.A., Sutherland, R.S., Kewley, L.J., 2008, ApJS, 178, 20
  \bibitem[\protect\citeauthoryear{Calderwood et al.}{2001}]{caldet01} Calderwood, T., Dobrzycki, A., Jessop, H., Harris, D.E.: 2001, ASP Conf. Ser., 238, 443 
  \bibitem[\protect\citeauthoryear{Cooke et al.}{2007}]{cooket07} Cooke, R., Kuncic, Z., Sharp, R., Bland-Hawthorn, J.: 2007, ApJ, 667, L163
  \bibitem[\protect\citeauthoryear{Cooke et al.}{2008}]{cooket08} Cooke, R., Bland-Hawthorn, J., Sharp, R., Kuncic, Z.: 2008, ApJ, 687, L29
  \bibitem[\protect\citeauthoryear{Cropper et al.}{2004}]{cropet04} Cropper, M., Soria, R., Mushotzky, R.F., Wu, K., Markwardt, C.B., Pakull, M.: 2004, MNRAS, 349, 39
  \bibitem[\protect\citeauthoryear{Dubner et al.}{1998}]{dubnet98}Dubner, G.M., Holdaway, M., Goss, W.M., Mirabel, I.F.: 1998, AJ, 116, 1842
  \bibitem[\protect\citeauthoryear{Erben et al.}{2005}]{erbeet05}Erben, T., et al.: 2005, AN, 326, 432
  \bibitem[\protect\citeauthoryear{Farrell et al.}{2009}]{farret09} Farrell, S.A., Webb, N.A., Barret, D., Godet, O., Rodrigues, J.M.: 2009, Nature, 460, 73
  \bibitem[\protect\citeauthoryear{Fender}{2001}]{fend01}Fender, R.P.: 2001, MNRAS, 322, 31
  \bibitem[\protect\citeauthoryear{Gallo et al.}{2005}]{gallet05}Gallo, E., Fender, R.P., Kaiser, C., Russell, D.M., Morganti, R., Oosterloo, T., Heinz, S.: 2005, Nature, 436, 819
  \bibitem[\protect\citeauthoryear{Gladstone}{2009}]{glad09} Gladstone, J.C.: 2009, Ph.D. Thesis
  \bibitem[\protect\citeauthoryear{Gladstone, Roberts \& Done}{Gladstone et al.}{2009}]{gladet09} Gladstone, J.C., Roberts, T.P., Done, C.: 2009, MNRAS, 397, 1836
  \bibitem[\protect\citeauthoryear{Goad et al.}{2002}]{goadet02} Goad, M.R., Roberts, T.P., Knigge, C., Lira, P.: 2002, MNRAS, 335, L67
  \bibitem[\protect\citeauthoryear{Hao \& Zhang}{2009}]{haozh09} Hao, J.F., Zhang, S.N.: 2009, ApJ, 702, 1648
  \bibitem[\protect\citeauthoryear{Heinz}{2002}]{hein02}Heinz, S.: 2002, A\&A, 388, L40
  \bibitem[\protect\citeauthoryear{Heinz et al.}{2008}]{heinet08}Heinz, S., Grimm, H.J., Sunyaev, R.A., Fender, R.P.: 2008, ApJ, 686, 1145
  \bibitem[\protect\citeauthoryear{Jin, Feng \& Kaaret}{Jin et al.}{2010}]{jinet10} Jin, J., Feng, H., Kaaret, P.: 2010, ApJ, 716, 181
  \bibitem[\protect\citeauthoryear{Kaaret, Ward, \& Zezas}{Kaaret et al.}{2004}]{kaaret04} Kaaret, P., Ward, M.J., Zezas, A.: 2004, MNRAS, 351, L83
  \bibitem[\protect\citeauthoryear{Kaaret \& Corbel}{2009}]{kaarco09} Kaaret, P., Corbel, S.: 2009, ApJ, 697, 950
  \bibitem[\protect\citeauthoryear{Kaiser et al.}{2004}]{kaiset04}Kaiser, C.R., Gunn, K.F., Brocksopp, C., Sokoloski, J.L.: 2004, ApJ, 612, 332
  \bibitem[\protect\citeauthoryear{K\"ording, Colbert \& Falcke}{K\"ording et al.}{2005}]{kordet05} K\"ording, E., Colbert, E., Falcke, H.: 2005, A\&A, 436, 427
  \bibitem[\protect\citeauthoryear{Liu \& Mirabel}{2005}]{liumi05} Liu, Q.Z., Mirabel, I.F.: 2005, A\&A, 429, 1125
  \bibitem[\protect\citeauthoryear{Lockman, Blundell \& Goss}{Lockman et al.}{2007}]{locket07} Lockman, F.J., Blundell, K.M., Goss, W.M.: 2007, MNRAS, 381, 881
  \bibitem[\protect\citeauthoryear{Miller-Jones et al.}{2008}]{millet08} Miller-Jones, J., Russell, D., Brocksopp, C., Sokoloski, J., Stappers, B., Muxlow, T.: 2008, AIPC, 1010, 50 (arXiv:0802.3446)
  \bibitem[\protect\citeauthoryear{Mirabel et al.}{1992}]{miraet92}Mirabel, I.F., Rodr\'{i}guez, L.F., Cordier, B., Paul, J., Lebrun, F.: 1992, Nature, 358, 215
  \bibitem[\protect\citeauthoryear{Pakull \& Mirioni}{2002}]{pakumi02} Pakull, M.W., Mirioni, L.: 2002, arXiv:astro-ph/0202488
  \bibitem[\protect\citeauthoryear{Pakull \& Gris\'e}{2008}]{pakugr08} Pakull, M.W., Gris\'e, F.: 2008, AIPC, 1010, 303 (arXiv:0803.4345)
  \bibitem[\protect\citeauthoryear{Pakull, Soria \& Motch}{Pakull et al.}{2010}]{pakuet10} Pakull, M.W., Soria, R., Motch, C.: 2010, Nature, 466, 209
  \bibitem[\protect\citeauthoryear{Ramsey et al.}{2006}]{ramset06} Ramsey, C.J., Williams, R.M., Gruendl, R.A., Chen, C.-H.R., Chu, Y.-H., Wang, Q.D.: 2006, ApJ, 641, 241
  \bibitem[\protect\citeauthoryear{Roberts et al.}{2001}]{robeet01} Roberts, T.P., Goad, M.R., Ward, M.J., Warwick, R.S., O'Brien, P.T., Lira, P., Hands, A.D.P.: 2001, MNRAS, 325, L7
  \bibitem[\protect\citeauthoryear{Roberts et al.}{2006}]{robeet06} Roberts, T.P., Kilgard, R.E., Warwick, R.S., Goad, M.R., Ward, M.J.: 2006, MNRAS, 371, 1877
  \bibitem[\protect\citeauthoryear{Russell et al.}{2007}]{russet07}Russell, D.M., Fender, R.P., Gallo, E., Kaiser, C.R., 2007, MNRAS, 376, 1341
  \bibitem[\protect\citeauthoryear{Sanders et al.}{2003}]{sandet03} Sanders, D.B., Mazzarella, J.M., Kim, D.-C., Surace, J.A., Soifer, B.T.: 2003, AJ, 126, 1607
  \bibitem[\protect\citeauthoryear{Sell et al.}{2010}]{sellet10} Sell, P.H., et al.: 2010: ApJ, 719, L194
  \bibitem[\protect\citeauthoryear{Soria et al.}{2005}]{soriet05} Soria, R., Cropper, M., Pakull, M., Mushotzky, R., Wu, K.: 2005, MNRAS, 356, 12
  \bibitem[\protect\citeauthoryear{Soria et al.}{2010}]{soriet10} Soria, R., Pakull, M.W., Broderick, J.W., Corbel, S., Motch, C.: 2010, MNRAS, in press (arXiv:1008.0394)
  \bibitem[\protect\citeauthoryear{Swartz, Tennant \& Soria}{Swartz et al.}{2009}]{swaret09} Swartz, D.A., Tennant, A.F., Soria, R.: 2009, ApJ, 703, 159
  \bibitem[\protect\citeauthoryear{Wiersema et al.}{2009}]{wieret09}Wiersema, K., et al.: 2009, MNRAS, 397, L6

\end{thebibliography}
\end{document}